\documentclass[aps,prl,reprint,letterpaper,twocolumn,showpacs,superscriptaddress]{revtex4}
\usepackage{graphicx}
\usepackage{xspace}
\usepackage{siunitx}
\usepackage{multirow}
\usepackage{lineno}

\newcommand{\met}{\ensuremath{\slash \kern-1.5ex E_T}\kern-.35ex\xspace}

\begin{document}


\title{Search for the Standard Model Higgs Boson Produced in Association with a $Z$ Boson in $p\bar{p}$ Collisions at $\sqrt{s} = 1.96$ TeV}
\affiliation{Institute of Physics, Academia Sinica, Taipei, Taiwan 11529, Republic of China}

\affiliation{Argonne National Laboratory, Argonne, Illinois 60439, USA}

\affiliation{University of Athens, 157 71 Athens, Greece}

\affiliation{Institut de Fisica d'Altes Energies, ICREA, Universitat Autonoma de Barcelona, E-08193, Bellaterra (Barcelona), Spain}

\affiliation{Baylor University, Waco, Texas 76798, USA}

\affiliation{Istituto Nazionale di Fisica Nucleare Bologna, $^{ee}$University of Bologna, I-40127 Bologna, Italy}

\affiliation{University of California, Davis, Davis, California 95616, USA}

\affiliation{University of California, Los Angeles, Los Angeles, California 90024, USA}

\affiliation{Instituto de Fisica de Cantabria, CSIC-University of Cantabria, 39005 Santander, Spain}

\affiliation{Carnegie Mellon University, Pittsburgh, Pennsylvania 15213, USA}

\affiliation{Enrico Fermi Institute, University of Chicago, Chicago, Illinois 60637, USA}

\affiliation{Comenius University, 842 48 Bratislava, Slovakia; Institute of Experimental Physics, 040 01 Kosice, Slovakia}

\affiliation{Joint Institute for Nuclear Research, RU-141980 Dubna, Russia}

\affiliation{Duke University, Durham, North Carolina 27708, USA}

\affiliation{Fermi National Accelerator Laboratory, Batavia, Illinois 60510, USA}

\affiliation{University of Florida, Gainesville, Florida 32611, USA}

\affiliation{Laboratori Nazionali di Frascati, Istituto Nazionale di Fisica Nucleare, I-00044 Frascati, Italy}

\affiliation{University of Geneva, CH-1211 Geneva 4, Switzerland}

\affiliation{Glasgow University, Glasgow G12 8QQ, United Kingdom}

\affiliation{Harvard University, Cambridge, Massachusetts 02138, USA}

\affiliation{Division of High Energy Physics, Department of Physics, University of Helsinki and Helsinki Institute of Physics, FIN-00014, Helsinki, Finland}

\affiliation{University of Illinois, Urbana, Illinois 61801, USA}

\affiliation{The Johns Hopkins University, Baltimore, Maryland 21218, USA}

\affiliation{Institut f\"{u}r Experimentelle Kernphysik, Karlsruhe Institute of Technology, D-76131 Karlsruhe, Germany}

\affiliation{Center for High Energy Physics: Kyungpook National University, Daegu 702-701, Korea; Seoul National University, Seoul 151-742, Korea; Sungkyunkwan University, Suwon 440-746, Korea; Korea Institute of Science and Technology Information, Daejeon 305-806, Korea; Chonnam National University, Gwangju 500-757, Korea; Chonbuk National University, Jeonju 561-756, Korea}

\affiliation{Ernest Orlando Lawrence Berkeley National Laboratory, Berkeley, California 94720, USA}

\affiliation{University of Liverpool, Liverpool L69 7ZE, United Kingdom}

\affiliation{University College London, London WC1E 6BT, United Kingdom}

\affiliation{Centro de Investigaciones Energeticas Medioambientales y Tecnologicas, E-28040 Madrid, Spain}

\affiliation{Massachusetts Institute of Technology, Cambridge, Massachusetts 02139, USA}

\affiliation{Institute of Particle Physics: McGill University, Montr\'{e}al, Qu\'{e}bec, Canada H3A~2T8; Simon Fraser University, Burnaby, British Columbia, Canada V5A~1S6; University of Toronto, Toronto, Ontario, Canada M5S~1A7; and TRIUMF, Vancouver, British Columbia, Canada V6T~2A3}

\affiliation{University of Michigan, Ann Arbor, Michigan 48109, USA}

\affiliation{Michigan State University, East Lansing, Michigan 48824, USA}

\affiliation{Institution for Theoretical and Experimental Physics, ITEP, Moscow 117259, Russia}

\affiliation{University of New Mexico, Albuquerque, New Mexico 87131, USA}

\affiliation{The Ohio State University, Columbus, Ohio 43210, USA}

\affiliation{Okayama University, Okayama 700-8530, Japan}

\affiliation{Osaka City University, Osaka 588, Japan}

\affiliation{University of Oxford, Oxford OX1 3RH, United Kingdom}

\affiliation{Istituto Nazionale di Fisica Nucleare, Sezione di Padova-Trento, $^{ff}$University of Padova, I-35131 Padova, Italy}

\affiliation{University of Pennsylvania, Philadelphia, Pennsylvania 19104, USA}

\affiliation{Istituto Nazionale di Fisica Nucleare Pisa, $^{gg}$University of Pisa, $^{hh}$University of Siena and $^{ii}$Scuola Normale Superiore, I-56127 Pisa, Italy}

\affiliation{University of Pittsburgh, Pittsburgh, Pennsylvania 15260, USA}

\affiliation{Purdue University, West Lafayette, Indiana 47907, USA}

\affiliation{University of Rochester, Rochester, New York 14627, USA}

\affiliation{The Rockefeller University, New York, New York 10065, USA}

\affiliation{Istituto Nazionale di Fisica Nucleare, Sezione di Roma 1, $^{jj}$Sapienza Universit\`{a} di Roma, I-00185 Roma, Italy}

\affiliation{Rutgers University, Piscataway, New Jersey 08855, USA}

\affiliation{Texas A\&M University, College Station, Texas 77843, USA}

\affiliation{Istituto Nazionale di Fisica Nucleare Trieste/Udine, I-34100 Trieste, $^{kk}$University of Udine, I-33100 Udine, Italy}

\affiliation{University of Tsukuba, Tsukuba, Ibaraki 305, Japan}

\affiliation{Tufts University, Medford, Massachusetts 02155, USA}

\affiliation{University of Virginia, Charlottesville, Virginia 22906, USA}

\affiliation{Waseda University, Tokyo 169, Japan}

\affiliation{Wayne State University, Detroit, Michigan 48201, USA}

\affiliation{University of Wisconsin, Madison, Wisconsin 53706, USA}

\affiliation{Yale University, New Haven, Connecticut 06520, USA}

\author{T.~Aaltonen}

\affiliation{Division of High Energy Physics, Department of Physics, University of Helsinki and Helsinki Institute of Physics, FIN-00014, Helsinki, Finland}

\author{B.~\'{A}lvarez~Gonz\'{a}lez$^z$}

\affiliation{Instituto de Fisica de Cantabria, CSIC-University of Cantabria, 39005 Santander, Spain}

\author{S.~Amerio}

\affiliation{Istituto Nazionale di Fisica Nucleare, Sezione di Padova-Trento, $^{ff}$University of Padova, I-35131 Padova, Italy}

\author{D.~Amidei}

\affiliation{University of Michigan, Ann Arbor, Michigan 48109, USA}

\author{A.~Anastassov$^x$}

\affiliation{Fermi National Accelerator Laboratory, Batavia, Illinois 60510, USA}

\author{A.~Annovi}

\affiliation{Laboratori Nazionali di Frascati, Istituto Nazionale di Fisica Nucleare, I-00044 Frascati, Italy}

\author{J.~Antos}

\affiliation{Comenius University, 842 48 Bratislava, Slovakia; Institute of Experimental Physics, 040 01 Kosice, Slovakia}

\author{G.~Apollinari}

\affiliation{Fermi National Accelerator Laboratory, Batavia, Illinois 60510, USA}

\author{J.A.~Appel}

\affiliation{Fermi National Accelerator Laboratory, Batavia, Illinois 60510, USA}

\author{T.~Arisawa}

\affiliation{Waseda University, Tokyo 169, Japan}

% ...

\collaboration{CDF Collaboration\footnote{With visitors from
$^a$Istituto Nazionale di Fisica Nucleare, Sezione di Cagliari, 09042 Monserrato (Cagliari), Italy,
$^b$University of CA Irvine, Irvine, CA 92697, USA,
$^c$University of CA Santa Barbara, Santa Barbara, CA 93106, USA,
$^d$University of CA Santa Cruz, Santa Cruz, CA 95064, USA,
$^e$Institute of Physics, Academy of Sciences of the Czech Republic, Czech Republic,
$^f$CERN, CH-1211 Geneva, Switzerland,
$^g$Cornell University, Ithaca, NY 14853, USA,
$^h$University of Cyprus, Nicosia CY-1678, Cyprus,
$^i$Office of Science, U.S. Department of Energy, Washington, DC 20585, USA,
$^j$University College Dublin, Dublin 4, Ireland,
$^k$ETH, 8092 Zurich, Switzerland,
$^l$University of Fukui, Fukui City, Fukui Prefecture, Japan 910-0017,
$^m$Universidad Iberoamericana, Mexico D.F., Mexico,
$^n$University of Iowa, Iowa City, IA 52242, USA,
$^o$Kinki University, Higashi-Osaka City, Japan 577-8502,
$^p$Kansas State University, Manhattan, KS 66506, USA,
$^q$Ewha Womans University, Seoul, 120-750, Korea,
$^r$University of Manchester, Manchester M13 9PL, United Kingdom,
$^s$Queen Mary, University of London, London, E1 4NS, United Kingdom,
$^t$University of Melbourne, Victoria 3010, Australia,
$^u$Muons, Inc., Batavia, IL 60510, USA,
$^v$Nagasaki Institute of Applied Science, Nagasaki, Japan,
$^w$National Research Nuclear University, Moscow, Russia,
$^x$Northwestern University, Evanston, IL 60208, USA,
$^y$University of Notre Dame, Notre Dame, IN 46556, USA,
$^z$Universidad de Oviedo, E-33007 Oviedo, Spain,
$^{aa}$CNRS-IN2P3, Paris, F-75205 France,
$^{bb}$Texas Tech University, Lubbock, TX 79609, USA,
$^{cc}$Universidad Tecnica Federico Santa Maria, 110v Valparaiso, Chile,
$^{dd}$Yarmouk University, Irbid 211-63, Jordan
}}

\noaffiliation



\date{\today}

\begin{abstract}
We present a search for the standard model Higgs boson produced in association with a $Z$ boson, using up to 7.9 fb$^{-1}$ of integrated luminosity from $p\bar{p}$ collisions collected with the CDF II detector.  We utilize several novel techniques, including multivariate lepton selection, multivariate trigger parametrization, and a multi-stage signal discriminant consisting of specialized functions trained to distinguish individual backgrounds.  By increasing acceptance and enhancing signal discrimination, these techniques have significantly improved the sensitivity of the analysis above what was expected from a larger dataset alone.  We observe no significant evidence for a signal, and we set limits on the $ZH$ production cross section.  For a Higgs boson with mass 115 GeV/$c^2$, we expect (observe) a limit of 3.9 (4.8) times the standard model predicted value, at the 95\% credibility level.  
\end{abstract}


\maketitle

The Higgs boson is the remaining unobserved particle of the standard model (SM) \cite{HiggsPW, Englert, Guralnik} predicted by the Higgs mechanism, which is postulated to describe the origin of electroweak symmetry breaking and elementary particle masses.  Direct searches at LEP and the Tevatron have excluded SM Higgs bosons with masses ($m_H$) below 114.4 GeV/$c^2$ \cite{LEPsearch} and in the range $156 \leq m_H \leq 177$ GeV/$c^2$ \cite{TevHighExcl}, respectively, at the 95\% credibility level (CL).  Recent results from the ATLAS and CMS experiments \cite{ATLASComb, CMSComb} have extended the excluded range of masses to $127 \leq m_H \leq 600$ GeV/$c^2$ (at the 95\% confidence level).

Production of Higgs bosons at the Tevatron primarily proceeds through the gluon fusion mechanism, $gg\to H$ \cite{HDECAY}. Low-mass Higgs bosons ($m_H < 135$ GeV/$c^2$) decay predominantly to a pair of $b$ quarks, with a branching fraction of 79\% (40\%) \cite{HDECAY} for $m_H = 100$ (135) GeV/$c^2$.  Due to overwhelming QCD multijet production,  low-mass searches with Higgs production via gluon fusion and $H\to b\bar{b}$ decay are not feasible.  To overcome this difficulty, we utilize the associated production of a Higgs boson with a massive vector boson, where leptonic decays of the vector boson produce distinctive event signatures.

This Letter presents a search for the SM Higgs boson using the $ZH\to\ell^+\ell^- b\bar{b}$ process, where $\ell$ is an electron ($e$) or muon ($\mu$).  We search for events containing two oppositely-charged leptons consistent with the decay of a $Z$ boson, and a hadronic signature consistent with the $H\to b\bar{b}$ decay mode. Previous searches \cite{CDFllbb, D0llbb} by the CDF and D0 collaborations have demonstrated that this final state provides good sensitivity to a Higgs boson signal, primarily due to the ability of the experiments to reconstruct both the $Z$ and Higgs bosons.  We study data from $p\bar{p}$ collisions at $\sqrt{s} = 1.96$ TeV recorded by the CDF II detector. We combine two independent analyses, one with $Z\to e^+e^-$ \cite{SarahThesis} and one with $Z\to \mu^+\mu^-$ \cite{JustinThesis}, using data corresponding to 7.5 and 7.9 fb$^{-1}$ of integrated luminosity, respectively.  

The CDF II detector \cite{Detector} consists of silicon-based and wire-drift-chamber tracking systems immersed in a 1.4 T magnetic field for particle momentum determination.  Surrounding the tracking systems are electromagnetic and hadronic calorimeters, providing coverage in the pseudorapidity \cite{coord} range $|\eta| < 3.6$.  Additional drift chambers used for muon identification are located in the outermost layer of the detector.

The sensitivity of this updated analysis is enhanced by using several novel techniques following two general strategies: increasing acceptance and enhancing signal discrimination.  To increase acceptance, we introduce artificial neural networks (NNs) for lepton selection, and we also use several online event-selection (trigger) algorithms not previously used.  Using a new technique, we are able to accurately model the combined behavior of these triggers, allowing access to $ZH$ candidate events beyond the reach of the previous CDF searches.  To enhance signal discrimination, we form a multi-stage event discriminant organized to isolate $ZH$ candidates from known SM and instrumental processes (backgrounds).

To improve on standard cut-based lepton identification, we instead select leptons consistent with the decay of a $Z$ boson by using several NNs.  Each NN identifies individual electrons or muons, distinguishing them from both non-leptonic candidates and true leptons not originating from $Z$ decays.  A single NN is used for muon identification, and is trained \cite{TMVA} to distinguish between true muons from simulated $Z$ decays and misidentified muons from a data sample containing same-charge muon candidates.  In events with $Z\to\mu\mu$ decays well contained in the detector, the muon NN selection achieves a $Z$ identification efficiency of $\sim$96\%, while simultaneously rejecting $\sim$94\% of the non-$Z$ background. Detector geometry \cite{CEM, PEM} motivates three NNs for electron identification. One is optimized for identification in the pseudorapidity range $|\eta| < 1.1$.  The other two NNs are trained for the forward regions; one considers only candidates with a silicon-based track and the other considers candidates without such a track in the region $1.1 \leq |\eta| \leq 2.8$.  Compared to the selection utilized in previous searches, the electron NN has improved the rejection of jets misidentified as electrons by a factor of five.  In total, the multivariate lepton selection has increased the acceptance of the analysis by $\sim$20\% over previous searches \cite{CDFllbb}.

Complementary to the improved lepton identification, we add additional triggers that were not previously utilized in this analysis channel.  Rather than using a single trigger with a threshold for muon $p_T$ or electron $E_T$ for the respective $Z$ selection, we consider any event selected by any trigger in three general sets.  The first set includes several triggers that select events containing muon detector and drift chamber activity indicative of a high-$p_T$ muon \cite{highPtMuon, xft}.  Included in this category may be triggers with lower $p_T$ thresholds than the default muon trigger.  The second set of triggers selects events with a large calorimeter-energy imbalance (missing transverse energy,\hspace{0.1cm} $\met$ \cite{MET}).  Some of these events contain muons that are not selected with the high-$p_T$ muon trigger, thereby increasing the acceptance of this analysis.  A third set of triggers selects events with activity in the calorimeter suggestive of a high-$E_T$ electron \cite{l2cal}.  By using these \emph{sets} of triggers rather than just single triggers for each lepton type, we increase the event selection acceptance by $\sim$10\%.  To model the complicated correlations between kinematic variables used in the trigger selection described above, we use a novel technique that uses NN functions to parametrize trigger efficiencies as a function of kinematic observables \cite{SarahThesis, JustinThesis}.

Utilizing the above strategies to increase acceptance, we select events containing opposite-sign \cite{forward_ele}, same-flavor lepton pairs with $m_{\ell\ell}$ in a window ($[76,106]$ GeV/$c^2$) centered on the mass of the $Z$ boson.  Additionally, we require at least two jets \cite{jets}, with transverse energy $E_T > 25$ GeV for the leading jet, and $E_T > 15$ GeV for all other jets.  All jets are required to come from the central region of the detector, $|\eta| < 2.0$.

We define a pre-tag region (PT) before applying $b$-quark jet identification, consisting of events with a reconstructed $Z$ boson and two or more jets passing the criteria described above.   We observe \num{33975} events in the PT region, and expect a total background yield of \num{34200} $\pm$ \num{4800} events, where the quoted uncertainty includes both systematic and statistical contributions. We expect 13.6 $\pm$ 1.1 $ZH$ signal events in the PT region, for $m_H = 115$ GeV/$c^2$.  The dominant process in the PT region is $Z$+light-flavor (LF) jets ($u$, $d$, $s$, and gluon jets), accounting for $\sim 85$\% of the total background.  $Z$+heavy-flavor (HF) ($b$ and $c$) jets  events, which contribute less than 10\% of the background, are a small contribution in the PT region, but become relatively more significant in the signal regions.  These processes are modeled using \textsc{alpgen} \cite{alpgen} to simulate the hard-scatter process, and \textsc{pythia} \cite{pythia} for the subsequent hadronization.  The $Z$+jets processes are simulated at leading order and require a $K$-factor of 1.4 \cite{Kfactor} for normalization to NLO cross sections. Other small backgrounds include diboson ($ZZ$, $WZ$, and $WW$) events and $t\bar{t}$ events, simulated entirely with \textsc{pythia} normalized to NLO \cite{VVNLO} and NNLO \cite{ttNNLO} predictions, respectively.  Finally, other processes, such as QCD multijet production, can produce two selected leptons in the event.   For muon events, this background is modeled using same-charge muon pairs from data.  For electron events, we measure the rate of jets passing the electron NN using collision data to estimate the contribution from these processes.  This background accounts for $\sim$3\% of the background in the PT region.  

We utilize two different $b$-quark-identifying algorithms to search for jets consistent with the $H\to b\bar{b}$ decay.  The secondary vertex algorithm (SV) \cite{SecVtx} identifies jets consistent with the decay of a long-lived $b$ hadron by searching for displaced vertices. The SV algorithm has both a tight and a loose operating point -- the loose point has better $b$-jet identification efficiency but also has a higher rate of jets incorrectly identified as $b$ jets.  The jet probability (JP) algorithm \cite{JetProb} uses track impact parameters relative to the primary vertex to construct a likelihood for all jet tracks to have originated from the primary vertex.  Both algorithms have imperfect rejection of $c$-quark jets, allowing some events containing them to contribute to the final signal regions.

We use the combination of the two highest-$E_T$ jets to form potential Higgs boson candidates.  We use a hierarchy of tag combinations to define three independent signal regions.  We first search for events with two tight SV tags -- defining the double-tag (DT) region, the most sensitive.  A second signal region includes events with one loose SV tag and one JP tag (LJP), and the third contains events with just one tight SV tag (ST).  These three regions are combined to search for $ZH$ production.  Table \ref{yields} shows the expected numbers of events for the signal and background processes, as well as the observed data.

\begin{table}
\caption{\label{yields}Expected background and observed data events for the three independent signal regions.  Also shown is the expected number of $ZH$ signal events, for a SM Higgs boson with $m_H = 115$ GeV/$c^2$. Quoted uncertainties include both systematic and statistical contributions.}
\begin{ruledtabular}
\begin{tabular}{lccc}

 \textbf{Process}  &  \textbf{ST} & \textbf{LJP} & \textbf{DT} \\ \hline
$Z$+LF, $Z+c\bar{c}$      &   683 $\pm$ 65 & 61 $\pm$ 9        & 7.6 $\pm$ 1.2    \\
$Z+b\bar{b}$   &     287 $\pm$ 72          & 58 $\pm$ 15       & 42 $\pm$ 10    \\

$t\bar{t}$       &    69 $\pm$ 7           & 29 $\pm$ 2        & 26 $\pm$ 3     \\
Diboson 			&    42 $\pm$ 3        & 9.5  $\pm$ 0.7    & 6.7 $\pm$ 0.6    \\
Other	 		 &   46 $\pm$ 12       & 3.4 $\pm$ 0.3     &  0.2 $\pm $ 0.1  \\
\hline
Background  	 &   1127 $\pm$ 134         & 160 $\pm$ 23     & 82 $\pm$ 15  \\
Data				&   1143 			    & 160              & 85    \\
$ZH$ (Predicted) & 4.5 $\pm$ 0.4             & 1.8 $\pm$ 0.1   & 1.7 $\pm$ 0.1 \\
 
\end{tabular}
\end{ruledtabular}
\end{table}

In this analysis, we use a one-dimensional signal discriminant while maintaining the simultaneous separation of $t\bar{t}$ and $Z+$jets events from the $ZH$ signal that was previously accomplished through a two-dimensional discriminant \cite{CDFllbb}. This method also further enhances signal discrimination by using two additional NNs in a multi-stage method, as described below.

We first train a NN signal discriminant, using several kinematic variables such as the dijet mass and \met, to distinguish the signal-like (trained with $ZH$ simulated events) and background-like (trained using a mixture of all background processes) events.  Each data and simulated event is sent through the same signal discriminants, with a unique function optimized for 11 different Higgs mass hypotheses, defined in increments of 5 GeV/$c^2$ between 100 and 150 GeV/$c^2$.
\begin{figure*}[t!]
\includegraphics[width=1.0\linewidth]{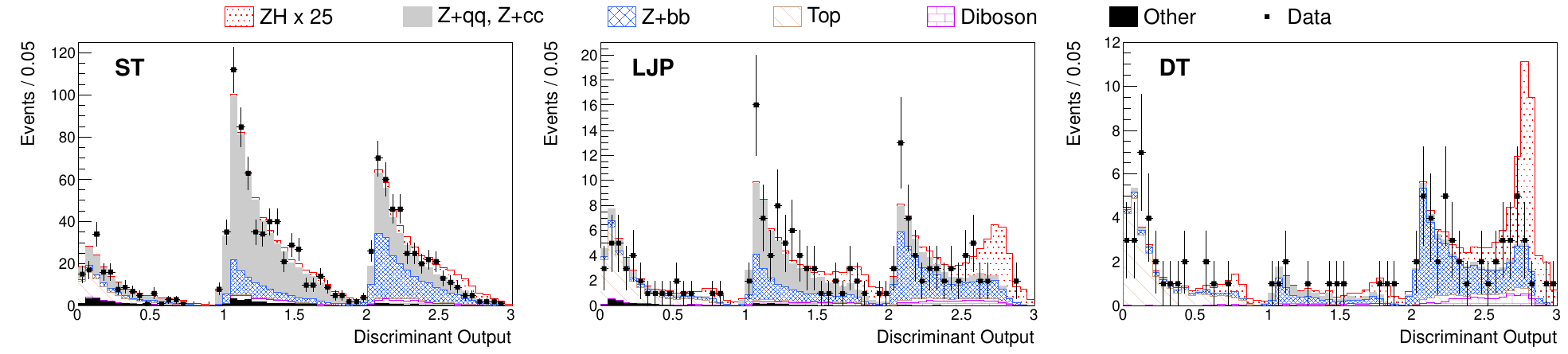}

\caption{Final discriminant output distributions for the three tag categories (ST, LJP, DT) used in this analysis.  The distributions shown are for the discriminant trained on $m_H = 115$ GeV/$c^2$ signal events.  The $ZH$ signal is shown, for $m_H$ = 115 GeV/$c^2$, and drawn scaled up by a factor of 25.}
\label{dists}
\end{figure*}

The multi-stage method defines three samples (I, II, III) where events can enter the final distributions used for limit setting.  The first step involves separating $t\bar{t}$ and $Z+$jets events.  This is done using a NN function trained to separate these specific processes.  A cut on the output of this discriminant is chosen to define a $t\bar{t}$-enhanced sample (Sample I).
Events which fail this cut and fall into Samples II or III are passed through a second NN function trained to separate $b$ jets from charm and light flavor jets \cite{KIT}.  A cut on the output of this flavor separator function defines a sample containing mainly $Z+c\bar{c}$ and $Z$+LF backgrounds (Sample II), and a region enriched in $b$ jets (Sample III).

This multi-stage approach produces final output distributions with three samples enriched in various background processes, as seen in Fig.\ \ref{dists}, where we add (0, 1, 2) to the signal discriminant output score for each event when the event falls in Sample (I, II, III) as described above.  By enhancing the signal discrimination in this way, we increase the sensitivity of the analysis by $\sim$10\% over the technique used in Ref. \cite{CDFllbb}.  We use these distributions to set limits on the $ZH$ production cross section times $H\to b\bar{b}$ branching ratio.

We evaluate several systematic uncertainties on the background and signal events.  A large source of systematic uncertainty arises from the cross section values used in the normalization of events: 40\%, 10\%, 6\%, and 5\%, for $Z+$HF \cite{Zbb_CDF}, $t\bar{t}$, diboson, and $ZH$ simulated events, respectively. An uncertainty of (1, 2, 5)\% is applied to the (ST, LJP, DT) $ZH$ samples after measuring changes in acceptance using simulated events with more or fewer particles radiated by the incoming and outgoing partons.  The mistag prediction is measured using data, and carries an uncertainty ranging from 13.5\% (ST) to 28.8\% (DT), depending on the tag category.  To account for differing $b$-jet identification efficiencies in data and simulated events, uncertainties of 5.2\% (ST), 8.7\% (LJP), and 10.4\% (DT) are applied to the $b$-tagged samples.  A 6\% uncertainty is applied to simulated events, accounting for uncertainty in the measurement of integrated luminosity.  The trigger model applied to simulated events requires a 5\% normalization uncertainty.   We also apply uncertainties on the lepton reconstruction efficiency and energy measurement of 1\% and 1.5\%, respectively.  For muons (electrons), we measure a 5\% (50\%) uncertainty on the normalization of the remaining background processes, based on differences in the rates of events containing same-charge and opposite-charge lepton pairs and in the rates of jets misidentified as electrons.  

In addition, we account for sources of uncertainty that also include shape variations to account for the migration of events in the final signal discriminant distributions when fluctuating these shape-defining quantities within their uncertainties.  These include uncertainties on the jet energies \cite{JetEnergyScale} as well as on the expected rate of $Z$ + mistag events.  

\begin{figure}
\includegraphics[width=1.0\columnwidth]{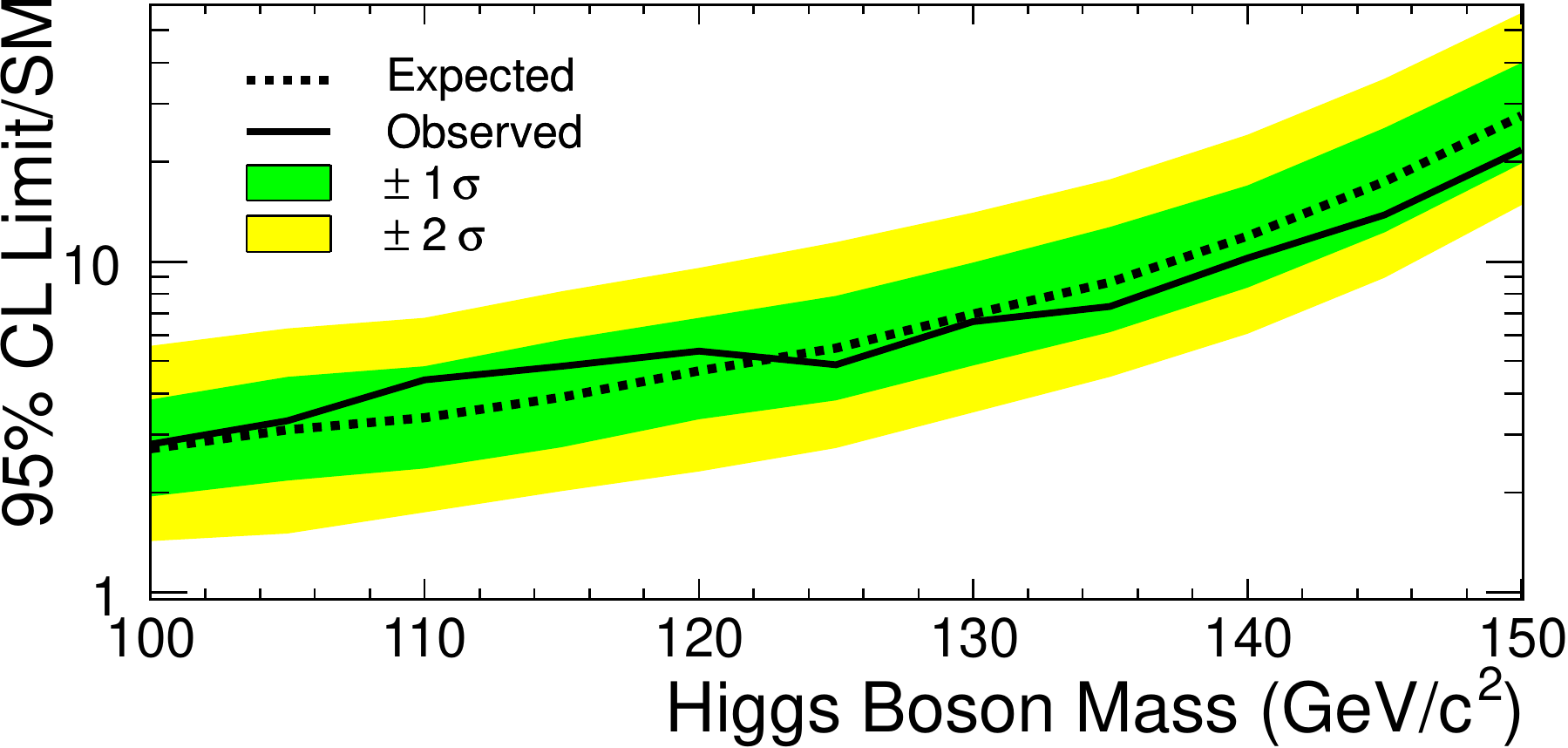}
\caption{Limits on the Higgs boson production cross section times the $H\to b\bar{b}$ branching ratio, given as a ratio to the standard model expected value.}
\label{limit_fig}
\end{figure}

Comparing the observed data to our background prediction including uncertainties, we do not find any evidence of a $ZH$ signal. We set upper limits on the $ZH$ production cross section times $H\to b\bar{b}$ branching ratio using a Bayesian algorithm \cite{PDG}, assuming a uniform prior on the signal rate.  We do this by performing simulated experiments, each with a pseudo-dataset generated by randomly varying the normalizations of background processes within their respective statistical and systematic uncertainties, taking into account all background expectations in the absence of a signal.  Each simulated experiment produces an upper limit on the $ZH$ production cross section.  The median of the 95\% CL upper limits from the simulated experiments is taken to be the expected 95\% CL upper limit of the analysis.  We define the 1-sigma and 2-sigma deviations on the expected limit as the bounds which contain 68.3\% and 95.5\%, respectively, of the simulated experiment results.  The observed data distribution is used to set the observed limit in a similar fashion. These limits are shown graphically, along with the 1-sigma and 2-sigma ranges, in Fig. \ref{limit_fig}.  We find that the observed limit is in good agreement with the expected limit for no signal, within the 1-sigma range across all Higgs mass hypotheses.

\begin{table}
\caption{\label{limits}Expected and observed 95\% CL limits on the $ZH$ production cross section times $H\to b\bar{b}$ branching ratio, relative to the expected standard model value, for each Higgs mass (in GeV/$c^2$) hypothesis.}
\begin{ruledtabular}
\begin{tabular}{c|ccccccccccc}
\boldmath{$m_H$} & \textbf{100} & \textbf{105}& \textbf{110}& \textbf{115}& \textbf{120}& \textbf{125}& \textbf{130}& \textbf{135}& \textbf{140}& \textbf{145}& \textbf{150} \\ \hline
Exp.& 2.7 & 3.1 & 3.4 & 3.9 & 4.7 & 5.5 & 7.0 & 8.7 & 12 & 17 & 28 \\
Obs.& 2.8 & 3.3 & 4.4 & 4.8 & 5.4 & 4.9 & 6.6 & 7.3 & 10 & 14 & 22 \\
\end{tabular}
\end{ruledtabular}
\end{table}

In conclusion, we have performed a search for the standard model Higgs boson in the process $ZH\to\ell^+\ell^- b\bar{b}$.  The sensitivity of this analysis has improved due to several new multivariate techniques, including multivariate  lepton identification, the use of NNs to obtain trigger efficiencies for simulated events, and a novel multi-stage discriminant approach used to enhance signal discrimination.  We observe no significant excess and set an upper limit on the $ZH$ production cross section times $H\to b\bar{b}$ branching ratio.  We expect (observe) a limit of 3.9 (4.8) times the standard model predicted value, for a Higgs boson with mass $m_H = 115$ GeV/$c^2$, at the 95\% CL.  The novel techniques presented here improve the sensitivity of the analysis by $\sim$25\% above the gain expected from the $\sim$85\% larger dataset.


We thank the Fermilab staff and the technical staffs of the participating institutions for their vital contributions. This work was supported by the U.S. Department of Energy and National Science Foundation; the Italian Istituto Nazionale di Fisica Nucleare; the Ministry of Education, Culture, Sports, Science and Technology of Japan; the Natural Sciences and Engineering Research Council of Canada; the National Science Council of the Republic of China; the Swiss National Science Foundation; the A.P. Sloan Foundation; the Bundesministerium f\"ur Bildung und Forschung, Germany; the Korean World Class University Program, the National Research Foundation of Korea; the Science and Technology Facilities Council and the Royal Society, UK; the Russian Foundation for Basic Research; the Ministerio de Ciencia e Innovaci\'{o}n, and Programa Consolider-Ingenio 2010, Spain; the Slovak R\&D Agency; the Academy of Finland; and the Australian Research Council (ARC). 
\bibliography{zhllbb}

\begin{thebibliography}{34}
\expandafter\ifx\csname natexlab\endcsname\relax\def\natexlab#1{#1}\fi
\expandafter\ifx\csname bibnamefont\endcsname\relax
  \def\bibnamefont#1{#1}\fi
\expandafter\ifx\csname bibfnamefont\endcsname\relax
  \def\bibfnamefont#1{#1}\fi
\expandafter\ifx\csname citenamefont\endcsname\relax
  \def\citenamefont#1{#1}\fi
\expandafter\ifx\csname url\endcsname\relax
  \def\url#1{\texttt{#1}}\fi
\expandafter\ifx\csname urlprefix\endcsname\relax\def\urlprefix{URL }\fi
\providecommand{\bibinfo}[2]{#2}
\providecommand{\eprint}[2][]{\url{#2}}

\bibitem[{\citenamefont{Higgs}(1964)}]{HiggsPW}
\bibinfo{author}{\bibfnamefont{P.~W.} \bibnamefont{Higgs}},
  \bibinfo{journal}{Phys. Rev. Lett.} \textbf{\bibinfo{volume}{13}},
  \bibinfo{pages}{508} (\bibinfo{year}{1964}).

\bibitem[{\citenamefont{Englert and Brout}(1964)}]{Englert}
\bibinfo{author}{\bibfnamefont{F.}~\bibnamefont{Englert}} \bibnamefont{and}
  \bibinfo{author}{\bibfnamefont{R.}~\bibnamefont{Brout}},
  \bibinfo{journal}{Phys. Rev. Lett.} \textbf{\bibinfo{volume}{13}},
  \bibinfo{pages}{321} (\bibinfo{year}{1964}).

\bibitem[{\citenamefont{Guralnik \emph{et~al}.}(1964)\citenamefont{Guralnik, Hagen,
  and Kibble}}]{Guralnik}
\bibinfo{author}{\bibfnamefont{G.~S.} \bibnamefont{Guralnik}},
  \bibinfo{author}{\bibfnamefont{C.~R.} \bibnamefont{Hagen}}, \bibnamefont{and}
  \bibinfo{author}{\bibfnamefont{T.~W.~B.} \bibnamefont{Kibble}},
  \bibinfo{journal}{Phys. Rev. Lett.} \textbf{\bibinfo{volume}{13}},
  \bibinfo{pages}{585} (\bibinfo{year}{1964}).

\bibitem[{\citenamefont{Barate \emph{et~al}.}(2003)}]{LEPsearch}
\bibinfo{author}{\bibfnamefont{R.}~\bibnamefont{Barate}} \bibnamefont{\emph{et~al}.}
  (\bibinfo{collaboration}{LEP Working Group}), \bibinfo{journal}{Phys. Lett.
  B} \textbf{\bibinfo{volume}{565}}, \bibinfo{pages}{61 }
  (\bibinfo{year}{2003}).

\bibitem[{\citenamefont{{The CDF and D0 Collaborations and the TEVNPH Working
  Group}}(2011)}]{TevHighExcl}
\bibinfo{author}{\bibnamefont{{The CDF and D0 Collaborations and the TEVNPH
  Working Group}}} (\bibinfo{year}{2011}), \eprint{arXiv:1107.5518,
  FERMILAB-CONF-11-354-E}.

\bibitem[{\citenamefont{Aad \emph{et~al}.}(2012)}]{ATLASComb}
\bibinfo{author}{\bibfnamefont{G.}~\bibnamefont{Aad}} \bibnamefont{\emph{et~al}.}
  (\bibinfo{collaboration}{{ATLAS Collaboration}}) (\bibinfo{year}{2012}),
  \eprint{{CERN-PH-EP-2012-019, arXiv:1202.1408}}.

\bibitem[{\citenamefont{Chatrchyan \emph{et~al}.}(2012)}]{CMSComb}
\bibinfo{author}{\bibfnamefont{S.}~\bibnamefont{Chatrchyan}}
  \bibnamefont{\emph{et~al}.} (\bibinfo{collaboration}{{CMS Collaboration}})
  (\bibinfo{year}{2012}), \eprint{{CERN-PH-EP-2012-023, arXiv:1202.1488}}.

\bibitem[{\citenamefont{Djouadi \emph{et~al}.}(1998)\citenamefont{Djouadi, Kalinowski,
  and Spira}}]{HDECAY}
\bibinfo{author}{\bibfnamefont{A.}~\bibnamefont{Djouadi}},
  \bibinfo{author}{\bibfnamefont{J.}~\bibnamefont{Kalinowski}},
  \bibnamefont{and} \bibinfo{author}{\bibfnamefont{M.}~\bibnamefont{Spira}},
  \bibinfo{journal}{Comput. Phys. Commun.} \textbf{\bibinfo{volume}{108}},
  \bibinfo{pages}{56 } (\bibinfo{year}{1998}).

\bibitem[{\citenamefont{Aaltonen \emph{et~al}.}(2010)}]{CDFllbb}
\bibinfo{author}{\bibfnamefont{T.}~\bibnamefont{Aaltonen}} \bibnamefont{\emph{et~al}.}
  (\bibinfo{collaboration}{CDF Collaboration}), \bibinfo{journal}{Phys. Rev.
  Lett.} \textbf{\bibinfo{volume}{105}}, \bibinfo{pages}{251802}
  (\bibinfo{year}{2010}).

\bibitem[{\citenamefont{Abazov \emph{et~al}.}(2010)}]{D0llbb}
\bibinfo{author}{\bibfnamefont{V.~M.} \bibnamefont{Abazov}}
  \bibnamefont{\emph{et~al}.} (\bibinfo{collaboration}{D0 Collaboration}),
  \bibinfo{journal}{Phys. Rev. Lett.} \textbf{\bibinfo{volume}{105}},
  \bibinfo{pages}{251801} (\bibinfo{year}{2010}).

\bibitem[{\citenamefont{Lockwitz}(2011)}]{SarahThesis}
\bibinfo{author}{\bibfnamefont{S.}~\bibnamefont{Lockwitz}}, Ph.D. thesis,
  \bibinfo{school}{Yale University} (\bibinfo{year}{2011}),
  \bibinfo{note}{{F}ERMILAB-THESIS-2012-02}.

\bibitem[{\citenamefont{Pilot}(2011)}]{JustinThesis}
\bibinfo{author}{\bibfnamefont{J.}~\bibnamefont{Pilot}}, Ph.D. thesis,
  \bibinfo{school}{The Ohio State University} (\bibinfo{year}{2011}),
  \bibinfo{note}{{F}ERMILAB-THESIS-2011-42}.

\bibitem[{\citenamefont{Acosta \emph{et~al}.}(2005{\natexlab{a}})}]{Detector}
\bibinfo{author}{\bibfnamefont{D.}~\bibnamefont{Acosta}} \bibnamefont{\emph{et~al}.}
  (\bibinfo{collaboration}{CDF Collaboration}), \bibinfo{journal}{Phys. Rev. D}
  \textbf{\bibinfo{volume}{71}}, \bibinfo{pages}{052003}
  (\bibinfo{year}{2005}{\natexlab{a}}).

\bibitem[{coo()}]{coord}
\bibinfo{note}{We use a cylindrical coordinate system with $z$ along the proton
  beam direction, $r$ the perpendicular radius from the central axis of the
  detector, and $\phi$ the azimuthal angle. For $\theta$ the polar angle from
  the proton beam, we define $\eta = -\ln\tan(\theta/2)$, 
  transverse momentum $p_T = p \sin\theta$ and transverse energy $E_T = E
  \sin\theta$.}

\bibitem[{\citenamefont{Hocker \emph{et~al}.}(2007)}]{TMVA}
\bibinfo{author}{\bibfnamefont{A.}~\bibnamefont{Hocker}} \bibnamefont{\emph{et~al}.},
  \bibinfo{journal}{PoS} \textbf{\bibinfo{volume}{ACAT}}, \bibinfo{pages}{040}
  (\bibinfo{year}{2007}).

\bibitem[{\citenamefont{Balka \emph{et~al}.}(1988)}]{CEM}
\bibinfo{author}{\bibfnamefont{L.}~\bibnamefont{Balka}} \bibnamefont{\emph{et~al}.},
  \bibinfo{journal}{Nucl. Instrum. Methods} \textbf{\bibinfo{volume}{267}},
  \bibinfo{pages}{272 } (\bibinfo{year}{1988}).

\bibitem[{\citenamefont{Albrow \emph{et~al}.}(2002)}]{PEM}
\bibinfo{author}{\bibfnamefont{M.}~\bibnamefont{Albrow}} \bibnamefont{\emph{et~al}.},
  \bibinfo{journal}{Nucl. Instrum. Methods} \textbf{\bibinfo{volume}{480}},
  \bibinfo{pages}{524 } (\bibinfo{year}{2002}).

\bibitem[{\citenamefont{Abulencia \emph{et~al}.}(2007)}]{highPtMuon}
\bibinfo{author}{\bibfnamefont{A.}~\bibnamefont{Abulencia}}
  \bibnamefont{\emph{et~al}.} (\bibinfo{collaboration}{CDF Collaboration}),
  \bibinfo{journal}{J. Phys. G} \textbf{\bibinfo{volume}{34}},
  \bibinfo{pages}{2457} (\bibinfo{year}{2007}).

\bibitem[{\citenamefont{Thomson \emph{et~al}.}(2002)}]{xft}
\bibinfo{author}{\bibfnamefont{E.}~\bibnamefont{Thomson}} \bibnamefont{\emph{et~al}.},
  \bibinfo{journal}{IEEE Trans. Nucl. Sci.} \textbf{\bibinfo{volume}{49}},
  \bibinfo{pages}{1063 } (\bibinfo{year}{2002}).

\bibitem[{MET()}]{MET}
\bibinfo{note}{The missing transverse energy, $\vec{\met}$, is defined by
  $\vec{\met} = - \sum_{i} E_T^i \hat{n}^i$, where $i$ is the number of the
  calorimeter tower with $|\eta| < 3.6$ and $\hat{n}^i$ is a unit vector
  perpendicular to the beam axis and pointing at the $i^{\mathrm{th}}$
  calorimeter tower. We also define $\met = |\vec{\met}|$.}

\bibitem[{\citenamefont{Bhatti \emph{et~al}.}(2009)}]{l2cal}
\bibinfo{author}{\bibfnamefont{A.}~\bibnamefont{Bhatti}} \bibnamefont{\emph{et~al}.},
  \bibinfo{journal}{IEEE Trans. Nucl. Sci.} \textbf{\bibinfo{volume}{56}},
  \bibinfo{pages}{1685 } (\bibinfo{year}{2009}).

\bibitem[{for()}]{forward_ele}
\bibinfo{note}{Forward ($|\eta| > 1.1$) electrons are exempt from this
  requirement}.

\bibitem[{\citenamefont{Blazey and Flaugher}(1999)}]{jets}
\bibinfo{author}{\bibfnamefont{G.~C.} \bibnamefont{Blazey}} \bibnamefont{and}
  \bibinfo{author}{\bibfnamefont{B.~L.} \bibnamefont{Flaugher}},
  \bibinfo{journal}{Annu. Rev. Nucl. Part. Sci.} \textbf{\bibinfo{volume}{49}},
  \bibinfo{pages}{633} (\bibinfo{year}{1999}).

\bibitem[{\citenamefont{Mangano \emph{et~al}.}(2003)\citenamefont{Mangano, Moretti,
  Piccinini, Pittau, and Polosa}}]{alpgen}
\bibinfo{author}{\bibfnamefont{M.~L.} \bibnamefont{Mangano}},
  \bibinfo{author}{\bibfnamefont{M.}~\bibnamefont{Moretti}},
  \bibinfo{author}{\bibfnamefont{F.}~\bibnamefont{Piccinini}},
  \bibinfo{author}{\bibfnamefont{R.}~\bibnamefont{Pittau}}, \bibnamefont{and}
  \bibinfo{author}{\bibfnamefont{A.~D.} \bibnamefont{Polosa}},
  \bibinfo{journal}{J. High Energy Phys.} \textbf{\bibinfo{volume}{07}},
  \bibinfo{pages}{001} (\bibinfo{year}{2003}).

\bibitem[{\citenamefont{Sjostrand \emph{et~al}.}(2006)\citenamefont{Sjostrand, Mrenna,
  and Skands}}]{pythia}
\bibinfo{author}{\bibfnamefont{T.}~\bibnamefont{Sjostrand}},
  \bibinfo{author}{\bibfnamefont{S.}~\bibnamefont{Mrenna}}, \bibnamefont{and}
  \bibinfo{author}{\bibfnamefont{P.~Z.} \bibnamefont{Skands}},
  \bibinfo{journal}{J. High Energy Phys.} \textbf{\bibinfo{volume}{05}},
  \bibinfo{pages}{026} (\bibinfo{year}{2006}).

\bibitem[{\citenamefont{Febres~Cordero
  \emph{et~al}.}(2008)\citenamefont{Febres~Cordero, Reina, and Wackeroth}}]{Kfactor}
\bibinfo{author}{\bibfnamefont{F.}~\bibnamefont{Febres~Cordero}},
  \bibinfo{author}{\bibfnamefont{L.}~\bibnamefont{Reina}}, \bibnamefont{and}
  \bibinfo{author}{\bibfnamefont{D.}~\bibnamefont{Wackeroth}},
  \bibinfo{journal}{Phys. Rev. D} \textbf{\bibinfo{volume}{78}},
  \bibinfo{pages}{074014} (\bibinfo{year}{2008}).

\bibitem[{\citenamefont{Campbell and Ellis}(1999)}]{VVNLO}
\bibinfo{author}{\bibfnamefont{J.~M.} \bibnamefont{Campbell}} \bibnamefont{and}
  \bibinfo{author}{\bibfnamefont{R.~K.}~\bibnamefont{Ellis}},
  \bibinfo{journal}{Phys. Rev. D} \textbf{\bibinfo{volume}{60}},
  \bibinfo{pages}{113006} (\bibinfo{year}{1999}).

\bibitem[{\citenamefont{Langenfeld \emph{et~al}.}(2009)\citenamefont{Langenfeld, Moch,
  and Uwer}}]{ttNNLO}
\bibinfo{author}{\bibfnamefont{U.}~\bibnamefont{Langenfeld}},
  \bibinfo{author}{\bibfnamefont{S.}~\bibnamefont{Moch}}, \bibnamefont{and}
  \bibinfo{author}{\bibfnamefont{P.}~\bibnamefont{Uwer}},
  \bibinfo{journal}{Phys. Rev. D} \textbf{\bibinfo{volume}{80}},
  \bibinfo{pages}{054009} (\bibinfo{year}{2009}).

\bibitem[{\citenamefont{Acosta \emph{et~al}.}(2005{\natexlab{b}})}]{SecVtx}
\bibinfo{author}{\bibfnamefont{D.}~\bibnamefont{Acosta}} \bibnamefont{\emph{et~al}.}
  (\bibinfo{collaboration}{CDF Collaboration}), \bibinfo{journal}{Phys. Rev. D}
  \textbf{\bibinfo{volume}{71}}, \bibinfo{pages}{052003}
  (\bibinfo{year}{2005}{\natexlab{b}}).

\bibitem[{\citenamefont{Abulencia \emph{et~al}.}(2006{\natexlab{a}})}]{JetProb}
\bibinfo{author}{\bibfnamefont{A.}~\bibnamefont{Abulencia}}
  \bibnamefont{\emph{et~al}.} (\bibinfo{collaboration}{CDF Collaboration}),
  \bibinfo{journal}{Phys. Rev. D} \textbf{\bibinfo{volume}{74}},
  \bibinfo{pages}{072006} (\bibinfo{year}{2006}{\natexlab{a}}).

\bibitem[{\citenamefont{Richter}(2007)}]{KIT}
\bibinfo{author}{\bibfnamefont{S.}~\bibnamefont{Richter}}, Ph.D. thesis,
  \bibinfo{school}{Karlsruhe Institute of Technology} (\bibinfo{year}{2007}),
  \bibinfo{note}{{F}ERMILAB-THESIS-2007-35}.

\bibitem[{\citenamefont{Abulencia \emph{et~al}.}(2006{\natexlab{b}})}]{Zbb_CDF}
\bibinfo{author}{\bibfnamefont{A.}~\bibnamefont{Abulencia}}
  \bibnamefont{\emph{et~al}.} (\bibinfo{collaboration}{CDF Collaboration}),
  \bibinfo{journal}{Phys. Rev. D} \textbf{\bibinfo{volume}{74}},
  \bibinfo{pages}{032008} (\bibinfo{year}{2006}{\natexlab{b}}).

\bibitem[{\citenamefont{Bhatti \emph{et~al}.}(2006)}]{JetEnergyScale}
\bibinfo{author}{\bibfnamefont{A.}~\bibnamefont{Bhatti}} \bibnamefont{\emph{et~al}.},
  \bibinfo{journal}{Nucl. Instrum. Methods A} \textbf{\bibinfo{volume}{566}},
  \bibinfo{pages}{375} (\bibinfo{year}{2006}).

\bibitem[{\citenamefont{Nakamura \emph{et~al}.}(2010)}]{PDG}
\bibinfo{author}{\bibfnamefont{K.}~\bibnamefont{Nakamura}} \bibnamefont{\emph{et~al}.}
  (\bibinfo{collaboration}{Particle Data Group}), \bibinfo{journal}{J. Phys. G}
  \textbf{\bibinfo{volume}{37}} (\bibinfo{year}{2010}).

\end{thebibliography}
\end{document}